\documentstyle[12pt]{article}

\author{Konstantin Kladko\\
Lifshitz Department of Theoretical Physics\\
Kharkov State University Kharkov 310077 Ukraine\\
\footnote{current address: MS-B258, LANL, 87545 USA}
e-mail kladko@lanl.gov}
\title{Einstein-Podolsky-Rosen paradox and  measurement of  quantum system.
}
\def\stackunder#1#2{\mathrel{\mathop{#2}\limits_{#1}}}

\begin{document}

\maketitle
\begin{abstract}
Einstein-Podolsky-Rosen (EPR) paradox is considered in a relation to a 
measurement of an arbitrary quantum system . It is shown that the EPR paradox
always appears in a gedanken experiment with two successively joined measuring
devices.

PACS\ number(s): 03.65
\end{abstract}

Einstein-Podolsky-Rosen paradox is a classical topic in quantum
mechanics(for elementary review see \cite{Bohm}). Here we show how
the EPR paradox takes place in a process of a measurement of a quantum
system by a measuring device, if the process of the measurement is
divided on two steps. Let us consider a gedanken setup, consisting of two
measuring devices $A$ and $B$, and an observer, watching the device $B$.
We assume  observables to take discrete,
nondegenerate values. The gedanken experiment consists of two steps.

1. Device $A$ measures an observable $F$ describing a bypassing elementary 
particle. Then some (large enough) period of time is waited in order
to assure that the particle is at a large distance from the measuring
devices.

2. Device $B$ measures the state of the device $A$ and the
observer observes the result of this measurement.

Below the behavior of  wave functions is considered.

1.  Initially the  
particle's wave function is $\Psi _p^{(0)}$ and the wave function of the device $A$
is $\Psi _A^{(0)}$,  the wave function of the  system as a whole is simply $%
\Psi _p^{(0)}\Psi _A^{(0)}$. After the step 1 
the wave function of the system as a whole is
$\stackunder{i,j}{\sum }C_{ij}\psi _i\varphi _j$, where $\psi _i$,$%
\varphi _j$ are eigenstates of the device $A$ and the particle
correspondingly. This  means that after the interaction with the
measuring device the particle is described not by a wave function, but by
a density matrix. Since the measuring device A  measures the observable $F$,
it means , by a definition, that the eigenstates of the
particle and the device $A$ have a one to one correspondence, and that the above some is
replaced by the sum $\sum a_i\psi _i\varphi _i$. Here $\psi _i$ is an
eigenfunction of the device, corresponding to the eigenfunction $\varphi _i$
with eigenvalue $F_i$ of the particle, $|$$a_i|^2$ is the probability for the
particle to have the value of $F$ equal to $F_i$ before the measurement.( see 
\cite{Landau} Chapter 1 \S 7 p.37).

2. During the second step the measuring device $B$ measures the state of the device 
$A$ and the observer observes the result of this measurement. As a result 
the device $A$ appears to be in some particular well-defined state $\psi _k$ . 
After this
second step the wave function of the system ''particle + device $A$''
becomes $\psi _k\varphi _k$, meaning that  measuring the state of the
device $A$  the state of the far away particle has been changed. 
The particle is now in  a definite state $\varphi _k$.  
This is the EPR paradox.

The result of the above consideration is: the EPR paradox always appears in the
process of measurement, if we subdivide the measurement into two stages. 
Therefore 
the EPR paradox can be found in any measurement, since any measuring
device we may  (in our minds) subdivide into two parts and there is 
no need to make 
special arrangements to observe the EPR paradox.

\end{document}